\documentclass[10pt]{article}
\usepackage[cp1251]{inputenc}
\usepackage[dvips]{epsfig}   \usepackage[dvips]{changebar}
\usepackage[dvips]{graphicx}

\def\lsim{\
  \lower-1.2pt\vbox{\hbox{\rlap{$<$}\lower5pt\vbox{\hbox{$\sim$}}}}\ }
\def\gsim{\
  \lower-1.2pt\vbox{\hbox{\rlap{$>$}\lower5pt\vbox{\hbox{$\sim$}}}}\ }

\usepackage{amsmath,enumerate, amsfonts, amssymb,amsthm}

\begin{document}
\title{Symmetry properties of the ground state \\ of the system of interacting spinless
bosons}
 \author{Maksim D. Tomchenko
\bigskip \\ {\small Bogolyubov Institute for Theoretical Physics} \\
 {\small 14b, Metrolohichna Str., Kyiv 03143, Ukraine}}
%\\
% {\small E-mail:mtomchenko@bitp.kiev.ua}}
 \date{\empty}
 \maketitle
 \large
 \sloppy
 \textit{We perform the symmetry analysis of the properties of the
ground state of a finite system of interacting spinless bosons for
the three most symmetric boundary conditions (BCs): zero BCs with
spherical and circular symmetries, as well as periodic BCs. The
symmetry of the system  can lead to interesting properties. For
instance, the density of a periodic Bose system is an exact
constant: $\rho(\textbf{r})=const$. Moreover, in the case of perfect
spherical symmetry of BCs, the crystalline state cannot produce the
Bragg peaks. The main result of the article is that symmetry
properties and general quantum-mechanical theorems admit equally
both crystalline and liquid ground state for a Bose system of any density. } 

 \section{Introduction}
As is known, it is impossible analytically and  very difficult
numerically to find directly and reliably the solution for the
ground state (GS) of a dense Bose system. The only exception is a
one-dimensional (1D) system of spinless point bosons: the analytical
structure of the ground-state wave function (WF) of such a system is
known \cite{ll1963,gaudin1971,gaudinm}. For all possible values of
the system parameters, the ground state of this system corresponds
to a liquid (gas) \cite{ll1963,mt2022}. For such a system,  gaseous
and liquid states are indistinguishable, see \cite{mt2022} for
details.

In what follows, we will consider symmetry properties of GS of a
system of interacting spinless bosons for several ideal boundary
conditions (BCs): zero BCs with rotational symmetry and periodic BCs
(such idealized BCs enables one to use corresponding mathematical
theorems). In these cases, the properties of the system are rather
bizarre and not visual. We will also try to clarify whether the
symmetry analysis allows us to find out the nature of the ground
state of a Bose system of a given density: is it a liquid or a
crystal?

The outline of the article is as follows. The general analysis of
the symmetry properties of one-dimensional, two-dimensional (2D),
and three-dimensional (3D) systems of  identical particles is made
in Sect.~2. Further, we investigate the symmetry properties of the
GS of a system of spinless bosons for the following cases: a
periodic 3D system (Sect.~3), a ball-shaped 3D system (Sect.~4), and
a perfectly circular 2D system (Sect.~5). In particular, in Sect.~5,
we find the structure of the wave functions of a 2D system of
spinless bosons with zero BCs on a circle. Some conclusions comprise
Sect.~6. In Appendix, we adduce the proof \cite{gilbert} of the
nondegeneracy of the GS of a 1D system of two spinless bosons and
point out conditions of validity of this proof.

\section{Symmetry analysis: general remarks}

Symmetry analysis enable us to draw some exact conclusions about
properties of the system. Below, we will investigate the symmetry
properties of the ground state of a system of interacting spinless
bosons under ideal BCs: periodic ones and zero BCs with rotational
symmetry. Such an analysis will allow us to comprehend the
properties of GS of real systems, for which BCs have lower symmetry.
We consider only spinless bosons since certain mathematical theorems
work just for such particles. Our analysis is valid for such inert
elements as $^{4}$He, Ne, Ar, Kr, and Xe.  For neon, argon, krypton,
and xenon, we have in mind isotopes corresponding to zero atomic
spin (consequently, zero nuclear spin, for atoms in the ground
state) and an infinite (or very long) lifetime: these are $^{20}$Ne,
$^{22}$Ne, $^{36}$Ar, $^{38}$Ar, $^{40}$Ar, $^{78}$Kr, $^{80}$Kr,
$^{82}$Kr, $^{84}$Kr, $^{86}$Kr, $^{124}$Xe, $^{126}$Xe, $^{128}$Xe,
$^{130}$Xe, $^{132}$Xe, $^{134}$Xe, and $^{136}$Xe
\cite{fastovskii1967,audi2003}.

It is noted in the book by J. Elliott and P. Dawber that GS of a
system should be invariant under all transformations of each of the
symmetry groups of the given system \cite{elliott} (in this case,
the system means the boundary-value problem: the Hamiltonian plus
BCs). In other words, GS should be the most symmetric state of the
system or enter the set of the most symmetric states.
% Here, the absence of spontaneous symmetry breaking is implicitly supposed.
Here it is implicitly assumed that there is no spontaneous symmetry
breaking.

Two theorems are valid for a system of interacting particles, which
is described by the Schr\"{o}dinger equation with zero BCs: (1) The
node theorem by R. Courant: if the eigenstates $\Psi_{j}$ are
numbered ($j=1,2,\ldots,\infty$) in the order of increasing energies
$E_{j}$, then the function $\Psi_{j}$ has no more than $j-1$ nodes
\cite{gilbert,courant}. (2) The theorem by R. Courant and D.
Hilbert: the ground state is non-degenerate \cite{gilbert}. These
theorems hold under the following conditions: all $\Psi_{j}$ are
one-component and real; the system is finite; the particles do not
have a spin or an intrinsic multipole moment; and the requirement
for the total interatomic potential: $-\infty < \int
d\textbf{r}_{1}\cdots
d\textbf{r}_{N}U(\textbf{r}_{1},\ldots,\textbf{r}_{N})|\Psi_{0}(\textbf{r}_{1},\ldots
,\textbf{r}_{N})|^{2}< \infty$ (see Appendix below). For a majority
of real-life systems, at least one of these conditions is violated.
However, all conditions hold for the atoms of inert elements
$^{4}$He, Ne, Ar, Kr, and Xe (for the interatomic potentials of
inert elements see \cite{aziz1984,aziz1991,rovenchak2000,mt2005}; we
do not consider hydrogen, because H$_{2}$ molecules possess
intrinsic quadrupole moment, which violates the conditions of the
theorems).

We note that these two theorems work for complex WFs as well. In the
sense that as the stationary Schr\"{o}dinger equation does not
contain complex numbers, the complete set of WFs can be constructed
so that all these WFs are real. The theorems have been proved for a
system of particles that possesses no definite (Bose or Fermi)
symmetry with respect to permutations: i.e. the complete set of WFs
contains the Bose symmetric, Fermi symmetric, and all other WFs.
However, the boundary-value problem is invariant under the
permutation group $S_{N}$. Therefore, the complete collection of WFs
can be set such that all WFs transform according to irreducible
representations of the group $S_{N}$ \cite{elliott,petrashen}. The
degeneracy multiplicity of the given state is equal to the dimension
of the irreducible representation the state transforms by
\cite{elliott,petrashen}. Therefore, a non-degenerate state
corresponds to a one-dimensional representation. As GS is
non-degenerate, the wave function of GS must transform according to
a one-dimensional representation. The group $S_{N}$ has exactly two
such representations: one corresponds to Bose symmetry, the other to
Fermi symmetry \cite{elliott,petrashen}. GS must have the highest
symmetry, so the ground-state WF must transform according to the
identity representation of the group $S_{N}$. Such a representation
corresponds to Bose symmetry. Therefore, those conclusions of the
theorems by Courant and Hilbert that concern the GS properties are
conclusions about the GS properties of a Bose system.

Note also that 1D and 2D systems of $N$ identical particles can have
special properties. In works \cite{leinaas1977,leinaas1991}, it was
made an analysis of the ``reduced'' configuration space
$M_{N}=(R^{N}-\Delta)/S_{N}$ obtained from the ordinary
configuration space $R^{N}$ of $N$ particles by identifying those
points from $R^{N}$, that differ only by permutation of coordinates
$\textbf{r}_{1}, \ldots, \textbf{r}_{N}$, and subtracting the set
$\Delta$ (this is a subspace of the space $R^{N}$, consisting of the
points for which the coordinates of two atoms coincide:
$\textbf{r}_{j}= \textbf{r}_{p}$ for any $j\neq p$). Let the space
$R$ have dimensionality $d$. From the standpoint of the properties
of the space $M_{N}$, the type of particle statistics is defined by
one-dimensional irreducible  representations of the first homotopy
group $\pi_{1}(M_{N})$ \cite{leinaas1977,khare}. For $d\geq 3$ we
have $\pi_{1}(M_{N})=S_{N}$, so for a 3D system ($d=3$) only Fermi
and Bose statistics are possible. However, for a 2D system ($d=2$),
the group $\pi_{1}(M_{N})$ is isomorphic to the infinite non-abelian
braid group $B_{N}$ \cite{khare,wu1984}. One-dimensional irreducible
representations of the group $B_{N}$ have the form
$\chi_{\theta}=e^{-i\theta}$, where  $\theta$ takes an infinite
number of values in the interval $[0,2\pi [$. The values $\theta=0$
and $\theta= \pi$ correspond to Bose and Fermi statistics,
respectively. The remaining values of $\theta$ correspond to
fractional (anomalous) statistics. In this case, fractional
statistics reveals itself both in the symmetry of WF (e.g.,
$\Psi(\textbf{r}_{1},\textbf{r}_{2})=e^{i\theta}\Psi(\textbf{r}_{2},\textbf{r}_{1})$
for $N=2$ [so called $\theta$ symmetry]; more precisely, the phase
$\theta$ arises when two particles are moved resulting in the
exchange \cite{khare}) and in the energy distribution of states and
thermodynamics \cite{khare}. We remark that anomalous statistics was
also introduced without using the symmetry properties of WFs
\cite{haldane1991,murthy1994a,wu1994a}.

WFs obeying fractional statistics are multivalued. These WFs usually
differ from each other by a constant factor that takes several
different values in the phase space. Such WFs describe several
different states of the same energy. It corresponds to a degenerate
state of the system.

Such anomalous WFs must be eigenfunctions for a given boundary-value
problem. Since the Hamiltonian of a 2D system of identical particles
is invariant under the group $S_{N}$ (rather than the more extensive
group $B_{N}$), only Bose and Fermi statistics are realized as a
rule. However, WFs with the $\theta$ symmetry do not correspond to
the $S_{N}$ symmetry of the Hamiltonian. Therefore, WFs satisfying
fractional statistics are possible only in the case of spontaneous
symmetry change. \textit{We suppose that spontaneous symmetry change
(or breakdown) occurs solely if at least one of the conditions of
Courant-Hilbert's theorem is violated.} Indeed, since the anomalous
states are degenerate, they cannot correspond to the genuine GS of
the system (according to Courant-Hilbert's theorem). Consequently,
if the GS is degenerate, then the conditions of the theorem are
violated. For characteristic problems leading to WFs with the
$\theta$ symmetry, the conditions of Courant-Hilbert's theorem are
actually violated (see Appendix below). Note that solutions
corresponding to anomalous statistics have been obtained
theoretically and justified experimentally (see monograph
\cite{khare} and reviews \cite{krive1987,laughlin1999}).

For 1D systems, a connection with anomalous statistics has also been
found. In particular, the energy distribution of states for a system
of spinless point bosons (potential
$U(x_{j}-x_{l})=2c\delta(x_{j}-x_{l})$) coincides with the
distribution for an ideal gas with generalized fractional statistics
\cite{bernard1994,isakov1994}, at any temperature $T$. For such a
system, WFs have Bose symmetry, and the unconventional statistics
manifests itself in the energy distribution of states and,
consequently, in thermodynamic formulae. However, the thermodynamic
formulae for such a system depend on the way of introducing
quasiparticles \cite{holes2020}. One can specify free quasiparticles
in such a way they to obey pure Bose statistics (for $N= \infty$,
$T\rightarrow 0$) \cite{mt2015}. Now let us imagine that we consider
two arbitrary quasiparticles, with momenta $k_{1}$ and $k_{2}$ say,
as \textit{one} quasiparticle with momentum $k_{3}=k_{1}+k_{2}$.
Accordingly, we consider two quasiparticles with momentum $k_{1}$
and two quasiparticles with momentum $k_{2}$ as two quasiparticles,
each having momentum $k_{3}=k_{1}+k_{2}$. And so on. We will obtain
another ensemble of quasiparticles, with a slightly different
distribution of quasiparticles on energies and with other occupation
numbers. Therefore, Bose statistics will be violated. Evidently, in
this way, one can obtain an infinite number of quasiparticle
ensembles with different statistics. In most cases, such an ensemble
will not obey Bose, Fermi, or fractional statistics. Thus, the
results in \cite{bernard1994,isakov1994,mt2015} show that for
$T\rightarrow 0$ there are ways of introducing quasiparticles in
which we arrive at Bose statistics or fractional statistics. And
just as importantly, the two approaches must be \textit{equivalent}:
experiments give only a single dependence of heat capacity on $T$,
which must be derived theoretically in both approaches. Indeed, the
approach with fractional statistics \cite{bernard1994,isakov1994} is
built on Yang-Yang's approach \cite{yangs1969} and thence is
equivalent to the latter. The Bose approach \cite{mt2015} also is
equivalent to Yang-Yang's approach, as shown in \cite{mt-therm}.
Consequently, the approaches \cite{bernard1994,isakov1994} and
\cite{mt2015} have indeed to be equivalent. For non-small $T$, Bose
quasiparticles are well-defined \cite{mtsp2019} but strongly
interact with each other, which greatly complicates the
thermodynamic description.
%(makes the thermodynamic description very complicated).
In this case, ``quasiparticles'' in the approach with fractional
statistics do not interact with each other even at large $T$, which
is a surprising and nontrivial property. It is natural to call
quasiparticles, corresponding to the most simple description of the
system, elementary excitations \cite{holes2020}. Therefore,
elementary excitations of the 1D system of spinless point bosons are
Bose quasiparticles at $T\rightarrow 0$ and quasiparticles,
satisfying generalized fractional statistics, at non-small $T$.
These properties are interesting. We also remark that the 1D
Calogero--Sutherland system of the particles interacting with a
$(x_{j}-x_{l})^{-2}$ potential can  be thermodynamically described,
at any temperature, as an ideal gas of (quasi) particles with
fractional statistics
\cite{bernard1994,isakov1994,polych1989,murthy1994b}.

Thus, 1D and 2D systems of identical particles can have nontrivial
statistical properties. However, \textit{below we consider only 2D
and 3D systems of identical particles for which the conditions of
Courant-Hilbert's theorem hold.  For such systems, there is no
spontaneous symmetry change, wave functions correspond to
irreducible representations of the group $S_{N}$, and GS is
described by a nodeless single-valued WF}. In that case, anomalous
statistics may in principle be obtained for an ``equivalent''
system, which can be matched to the given system similarly to the
analysis in \cite{bernard1994,isakov1994,polych1989,murthy1994b}.
Below we are interested in the properties of an initial real-life
system. As is seen from the aforesaid,  the symmetry of the
ground-state WFs of such a system must correspond to Bose
statistics.

% Presumably, for all systems where spontaneous symmetry breaking is
%observed and GS is degenerate, at least one of the conditions of
%Courant-Hilbert's theorems is violated.

So, GS of a system of spinless bosons must be described by a
nodeless non-degenerate WF belonging to the set of functions that
are invariant under all transformations of symmetry groups of the
boundary-value problem. Let us consider the properties of wave
functions for several boundary-value problems and check whether 
crystalline and liquid GS meet these requirements.

Since the symmetry of the boundary-value problem means the symmetry
of both the Hamiltonian and the boundary conditions, we will
consider idealized BCs corresponding to symmetries of the
Hamiltonian: (i) periodic BCs and (ii) zero ones with rotational
symmetry (3D ball and 2D disk).

We will consider only finite systems (since for an infinite system
Courant-Hilbert's theorem does not work, generally speaking, and
spontaneous symmetry breaking is possible).

\section{Periodic 3D system of spinless bosons}
The Hamiltonian
\begin{equation}
 \hat{H} = -\frac{\hbar^{2}}{2m}\sum\limits_{j=1}^{N}\triangle_{\textbf{r}_{j}} + \frac{1}{2}\mathop{\sum_{j,l}}\limits_{j\neq l}
 U(|\textbf{r}_{l}-\textbf{r}_{j}|)
     \label{2-1} \end{equation}
is invariant under the continuous group of translations $T(3)$,  the
continuous group of rotations $SO(3)$, the group of permutations
$S_{N}$, and the group of inversions $C_{i}$ (the latter contains
two elements: inversion $I$ and $I^{2}=1$; here and below, we
consider that any symmetry transformation applies to
% is performed with
all coordinates $\textbf{r}_{1},\ldots,\textbf{r}_{N}$). Since the
space is isotropic and uniform, the invariance of $\hat{H}$ under
the groups $SO(3)$ and $T(3)$ would survive if $\hat{H}$ also
includes three-particle interaction
\cite{axilrod1943,bruch1973,loubeyre1988,boronat1994} and possible
many-particle ones. Periodic BCs are invariant under the groups
$T(3)$, $S_{N}$,  and $C_{i}$, but are not under the group $SO(3)$.
The ground-state WF reads
\begin{equation}
\Psi_{0} =  e^{S_{0}}, \label{2-2}    \end{equation}
\begin{eqnarray}
S_{0}(\textbf{r}_{1},\ldots,\textbf{r}_{N}) &=&
\mathop{\sum_{j_{1}j_{2}}}\limits_{j_{1}< j_{2}}
S_{2}(\textbf{r}_{j_{1}}-\textbf{r}_{j_{2}})+
\mathop{\sum_{j_{1}j_{2}j_{3}}}\limits_{j_{1}<
j_{2}<j_{3}}S_{3}(\textbf{r}_{j_{1}}-\textbf{r}_{j_{2}},\textbf{r}_{j_{2}}-\textbf{r}_{j_{3}})+\ldots
\nonumber \\ &+&\mathop{\sum_{j_{1}j_{2}\ldots
j_{N}}}\limits_{j_{1}< j_{2}<\ldots
<j_{N}}S_{N}(\textbf{r}_{j_{1}}-\textbf{r}_{j_{2}},\textbf{r}_{j_{2}}-\textbf{r}_{j_{3}},\ldots
, \textbf{r}_{j_{N-1}}-\textbf{r}_{j_{N}}).
 \label{2-3}    \end{eqnarray}
Such a WF can describe both a liquid
\cite{holes2020,gross1962,woo1972,feenberg1974,yuv1} and a crystal
\cite{mcmillan1965,chester1970,reatto1995,whitlock2006}. This WF is
invariant under the groups $T(3)$, $S_{N}$, and $C_{i}$. The
transition from the liquid GS to the crystalline one leads to the
appearance of long-range crystalline order in the two-particle and
higher correlation functions. \textit{Thus, the properties of the
crystalline and liquid GS are in agreement with the theorems by
Courant and Hilbert and with Elliott-Dawber's statement.}

We note that any state of the system with the total momentum
$\textbf{P}\neq 0$ must be degenerate because
$[\hat{H},\hat{\textbf{P}}]=0$, $[\hat{H},\hat{I}_{r}]=0$, but
$[\hat{\textbf{P}},\hat{I}_{r}]=-2\hat{I}_{r}\hat{\textbf{P}}\neq 0$
(here, $\hat{I}_{r}$ denotes the coordinate inversion operator:
$\hat{I}_{r}\Psi(\textbf{r}_{1},\ldots,\textbf{r}_{N})=\Psi(-\textbf{r}_{1},\ldots,-\textbf{r}_{N})$).
On states with zero $\textbf{P}$ the operators $\hat{\textbf{P}}$
and $\hat{I}_{r}$ commute with each other. Only such states can be
non-degenerate.  The condition $\textbf{P}=0$ is satisfied for the
ground state (\ref{2-2}), (\ref{2-3}) and an infinite number of
excited states. All such states are invariant under the groups
$T(3)$, $S_{N}$, and $C_{i}$. These are the most symmetric states of
the system.

The ground state of a Bose crystal is most often described by the
localized ansatz \cite{nosanow1966,guyer}
 \begin{equation}
   \Psi^{c}_{0} \approx  e^{S_{0}}\sum\limits_{P_{c}}\prod\limits_{j=1}^{N} \varphi(\textbf{r}_{j}-\textbf{R}_{j}),
 \label{crys-1}    \end{equation}
where $\textbf{r}_{j}$ and $\textbf{R}_{j}$ are the coordinates of
atoms and lattice sites, respectively, $N$ is the number of atoms in
the system, $\varphi(\textbf{r})=e^{- \alpha r^{2}/\bar{d}^{2}}$
($\bar{d}$ stands for the average interatomic distance), $P_{c}$
denotes all possible permutations of coordinates $\textbf{r}_{j}$,
and $e^{S_{0}}$ corresponds to the isotropic liquid solution with
$S_{0}$ (\ref{2-3}).
%Here, $\textbf{R}_j$ are fixed and are the same for all
%possible configurations $\{\textbf{r}_{j}\}$.
Such ansatz can be considered as an approximate solution under zero
BCs. It gives satisfactory agreement with experiments
\cite{nosanow1966,cazorla2008}.

Interestingly, for any pure state (including GS) of a periodic Bose
system,   the concentration is an exact constant:
$n(\textbf{r})=const$ \cite{vak1989,vak1990,sacha2018,mt2020}. This
property is related to translation invariance and can be proved
easily \cite{sacha2018,mt2020}. For a system with $T>0,$ this can be
proved analogously to the analysis in \cite{sacha2018}, using the
formula $n(\textbf{r})=const\cdot\int d\textbf{r}_{2}\ldots
d\textbf{r}_{N}\sum_{j}e^{-E_{j}/k_{B}T}
|\Psi_{j}(\textbf{r},\textbf{r}_{2},\ldots,\textbf{r}_{N})|^{2}$ and
property $\Psi(\textbf{r}_{1}+\delta
\textbf{r},\ldots,\textbf{r}_{N}+\delta \textbf{r}) = (1+i
\textbf{p}\delta
\textbf{r})\Psi(\textbf{r}_{1},\ldots,\textbf{r}_{N})= e^{i
\textbf{p}\delta
\textbf{r}}\Psi(\textbf{r}_{1},\ldots,\textbf{r}_{N})$. In this
case, the crystalline ordering is hidden: it manifests itself in
long-range oscillations (with periods  of a crystal) of the
two-particle and higher distribution functions, not in the density.

As an example, consider  $N$ free spinless bosons in the state
$|\Psi^{c}\rangle =
|N_{\textbf{k}_{x}},N_{-\textbf{k}_{x}},N_{\textbf{k}_{y}},N_{-\textbf{k}_{y}},N_{\textbf{k}_{z}},N_{-\textbf{k}_{z}},N_{\textbf{k}=0}\rangle$,
where $N_{\textbf{k}}$ is the number of Bose particles with momentum
$\hbar\textbf{k}$. Let us set
$N_{\textbf{k}_{x}}=N_{-\textbf{k}_{x}}=N_{\textbf{k}_{y}}=N_{-\textbf{k}_{y}}=N_{\textbf{k}_{z}}=N_{-\textbf{k}_{z}}=J$
and $N_{\textbf{k}=0}=N-6J$. Such a state is described by the second
quantized operator
 \begin{eqnarray}
\hat{\psi}(\textbf{r},t) &=&V^{-1/2}\left
(\hat{a}_{\textbf{k}_{x}}e^{ik_{x}x}+\hat{a}_{-\textbf{k}_{x}}e^{-ik_{x}x}+
\hat{a}_{\textbf{k}_{y}}e^{ik_{y}y}\right.\nonumber \\ &+& \left.
\hat{a}_{-\textbf{k}_{y}}e^{-ik_{y}y}+
\hat{a}_{\textbf{k}_{z}}e^{ik_{z}z}+\hat{a}_{-\textbf{k}_{z}}e^{-ik_{z}z}+\hat{a}_{0}\right
).
       \label{Psi-n}\end{eqnarray}
After some evaluation, we find the two-particle distribution
function:
 \begin{eqnarray}
  && g_{2}(\textbf{r}_{1},\textbf{r}_{2}) \equiv  C_{g}\langle \Psi_{0}^{c}|\hat{\psi}^{+}(\textbf{r}_{1},t)\hat{\psi}^{+}(\textbf{r}_{2},t)
\hat{\psi}(\textbf{r}_{1},t)\hat{\psi}(\textbf{r}_{2},t)|\Psi_{0}^{c}\rangle
 \nonumber \\ && =
 C_{g}V^{-2}\left [N^{2}-N-6J^{2}+4JF_{3}\cdot(JF_{3}+N-6J)\right ],
       \label{g2-1}\end{eqnarray}
where
 \begin{eqnarray}
 F_{3}=\cos{[k_{x}(x_{1}-x_{2})]}+\cos{[k_{y}(y_{1}-y_{2})]}+\cos{[k_{z}(z_{1}-z_{2})]},
       \label{Psi-n}\end{eqnarray}
$C_{g}=V^{2}/(N^{2}-N)$. Here, the normalization is $\int
d\textbf{r}_{1}d\textbf{r}_{2}g_{2}(\textbf{r}_{1},\textbf{r}_{2})=V^{2}$.
It is a translationally invariant crystal-like solution with a
rectangular 3D lattice. The amplitude of oscillations of the
function $g_{2}(\textbf{r}_{1},\textbf{r}_{2})$  is greatest at
$J=N/6$, i.e. when $N_{\textbf{k}=0}=0$. Such a state corresponds to
WF with many nodes. In this case, GS corresponds obviously to the
state $|\Psi_{0}\rangle = |N_{\textbf{k}=0}\rangle$ with
$N_{\textbf{k}=0}=N$. We may expect that, at \textit{nonzero}
interatomic interaction, the state $|\Psi^{c}\rangle =
|N_{\textbf{k}_{x}},N_{-\textbf{k}_{x}},N_{\textbf{k}_{y}},N_{-\textbf{k}_{y}},N_{\textbf{k}_{z}},N_{-\textbf{k}_{z}},N_{\textbf{k}=0}\rangle$
(where $N_{\textbf{k}}$ is the number of \textit{phonons} with
momentum $\hbar\textbf{k}$) also corresponds to a crystal with a
nodal WF. The exact crystalline solution \cite{mt2022} for a 1D
system of point bosons indicates it (this solution has the same
structure). Such a crystal can be regarded as a liquid with a
condensate of phonons (this interpretation is correct at least for
the 1D system of point bosons). In this case,  the condensate of
phonons creates a lattice. On the other hand, the lattice forms by a
network of nodes of WF (the exact solution in \cite{mt2022} also has
this property). One can expect that the condensate of phonons with
momentum $\hbar\textbf{k}$ goes along with a condensate of atoms
with the same momentum.

We remark that crystalline solutions with a condensate of atoms,
that possess nonzero momentum  (quasimomentum), were considered in a
number of works
\cite{mt2020,gross1958,gross1960,luca,coniglio1969,kirz,nep,shlyapa2015,andreev2017,fil2020}.
Strictly speaking, all of these solutions may not correspond to the
genuine GS with a nodeless WF. D. Kirzhnits and Yu.~Nepomnyashchii
have called such a solution by ``coherent crystal'' \cite{kirz,nep}.
The energy of a state with a multi-node WF considerably exceeds the
GS energy of the system. It is yet unclear, whether such coherent
crystal can be stable and exist in Nature.

\section{Ball-shaped 3D system of spinless bosons} Consider a 3D
system of spinless bosons placed in a sphere with zero BCs ($\Psi=0$
on the boundary). Such a system has the form of an ideal ball. The
Hamiltonian $\hat{H}$ (\ref{2-2}) is invariant under rotations.
Therefore, $\hat{H}$ commutes with the operators of the total
angular momentum of the system $\hat{L}_{x}$, $\hat{L}_{y}$,
$\hat{L}_{z},$ and $\hat{\textbf{L}}^{2}$
\cite{petrashen,land3,vak}. BCs are also invariant under rotations.
Hence, the complete system of wave functions for such a
boundary-value problem can be constructed so that they are the
eigenfunctions of the operators $\hat{H}$, $\hat{\textbf{L}}^{2},$
and $\hat{L}_{z}$ \cite{land3,vak}. In this case, each WF satisfies
the relation $\hat{\textbf{L}}^{2}\Psi_{L}=\hbar^{2}
L(L+1)\Psi_{L}$, where $L=0, 1, 2,\ldots, \infty$. The states with
$L\neq 0$ are $(2L+1)$-fold degenerate due to the noncommutativity
of the operators $\hat{L}_{x}$, $\hat{L}_{y}$, and $\hat{L}_{z}$
\cite{land3}. Solely the states with $L= 0$ are non-degenerate.

The last statement can be proved in the following more general way.
$\hat{H}$ is invariant under rotations
$\textbf{r}_{j}\rightarrow\acute{\textbf{r}}_{j}=A\textbf{r}_{j}$
\cite{land3,vak} (here, $j=1,\ldots,N$,  $A$ is the rotation matrix,
the vectors $\textbf{r}_{j}$ and $\acute{\textbf{r}}_{j}$ are given
in the same basis, and the coordinate origin is in the center of the
ball). Therefore, $\hat{H}$ commutes with the rotation operator
$\hat{R}=e^{i\varphi \textbf{i}_{\varphi}\hat{\textbf{L}}/\hbar}$
\cite{petrashen,land3,vak}, where $\varphi$ is a rotation angle, and
the unit vector $\textbf{i}_{\varphi}$ sets the rotation axis.
Hence, $[\hat{H},\hat{\textbf{L}}]=0,$ and
$[\hat{H},\hat{\textbf{L}}^{2}]=0$. Since the Hamiltonian and BCs
are invariant under the rotations, WFs can be set so that they
transform by irreducible representations of the rotation group
$SO(3)$ \cite{elliott,petrashen} (this is true for unitary
representations; as the group $SO(3)$ is compact, its
representations are equivalent to unitary ones
\cite{petrashen,gelfand}). This group is characterized by the
complete collection of irreducible representations $g\rightarrow
\hat{T}_{l}(g)$ with $l=0, 1/2, 1, 3/2, 2, \ldots, \infty $ and the
dimension $2l+1$ (here, $g$ is an element of the group). The
representations with integer and half-integer $l$ are, respectively,
one- and two-valued \cite{elliott,petrashen,gelfand,golod}. The
scalar WFs can transform only by the representations with integer
$l$. In this case, $\hat{\textbf{L}}^{2}\Psi^{(l)}=\hbar^{2}
l(l+1)\Psi^{(l)}$ \cite{elliott,petrashen} (in this article, we do
not  distinguish between the numbers $L$ and $ l$). The rotation
operator is defined by the formula \cite{land3,vak}
\begin{equation}
 \hat{R}\Psi(\textbf{r}_{1},\ldots,\textbf{r}_{N}) = \Psi(\acute{\textbf{r}}_{1},\ldots,\acute{\textbf{r}}_{N}),
     \label{1-new0} \end{equation}
where $\textbf{r}_{j}$ and $\acute{\textbf{r}}_{j}=A\textbf{r}_{j}$
are the coordinates of a vector before and after a rotation.  Eq.
(\ref{1-new0}) leads  to the formula $\hat{R}=e^{i\varphi
\textbf{i}_{\varphi}\hat{\textbf{L}}/\hbar}$ \cite{land3,vak}. The
wave functions, invariable under any rotation ($\hat{R}\Psi \equiv
e^{i\varphi \textbf{i}_{\varphi}\hat{\textbf{L}}/\hbar}\Psi =\Psi$,
i.e., $\hat{\textbf{L}}\Psi=0$), transform according to the identity
representation $g\rightarrow\hat{T}_{0}(g)$:
$\hat{T}_{0}(g)\Psi=\Psi$ for any element  $g$ of the group $SO(3)$.
The irreducible representation $g\rightarrow\hat{T}_{l}(g)$ is
characterized by the orthonormal basis
$\Psi^{(l)}_{1},\Psi^{(l)}_{2},\ldots,\Psi^{(l)}_{2l+1}$. In this
case,
$\hat{T}_{l}(g)\Psi^{(l)}_{j}=\sum_{p=1}^{2l+1}T^{(l)}_{pj}(g)\Psi^{(l)}_{p}$
for any element  $g$ of the group $SO(3)$, where $T^{(l)}(g)$ are
the matrices of constants and realize the representation
$g\rightarrow\hat{T}_{l}(g)$ \cite{elliott,petrashen}. For each
representation $g\rightarrow\hat{T}_{l}(g),$ all functions
$\Psi^{(l)}_{p}$ correspond to the same energy. Indeed, let
$\Psi^{(l)}_{j}$ be an eigenfunction of the Schr\"{o}dinger equation
with energy $E$:
\begin{equation}
 \hat{H}\Psi^{(l)}_{j} = E\Psi^{(l)}_{j}.
     \label{1-new00} \end{equation}
Let us act by the operator
$\hat{T}(g)=\hat{R}^{-1}(g)=\hat{R}(g^{-1})$ \cite{elliott,gelfand}
on this equation. Since $\hat{R}(g)\hat{H}-\hat{H}\hat{R}(g)=0$ for
any rotation $g$, the last equality holds also for the rotation
$g^{-1}$. From whence, we get
$\hat{T}(g)\hat{H}-\hat{H}\hat{T}(g)=0$. Therefore,
\begin{equation}
 E\hat{T}(g)\Psi^{(l)}_{j}=\hat{T}(g)E\Psi^{(l)}_{j}=\hat{T}(g)\hat{H}\Psi^{(l)}_{j}
=\hat{H}\hat{T}(g)\Psi^{(l)}_{j}.
     \label{1-new000} \end{equation}
That is, the function $\hat{T}(g)\Psi^{(l)}_{j}$ is also an
eigenfunction of the Schr\"{o}dinger equation with energy $E$. We
now substitute the expansion $\hat{T}(g)\Psi^{(l)}_{j}\equiv
[\hat{T}_{0}(g)\oplus
\hat{T}_{1}(g)\oplus\ldots\oplus\hat{T}_{\infty}(g)]\Psi^{(l)}_{j}=\hat{T}_{l}(g)\Psi^{(l)}_{j}=\sum_{p=1}^{2l+1}T^{(l)}_{pj}(g)\Psi^{(l)}_{p}$
in formula (\ref{1-new000}). Since the basis functions
$\Psi^{(l)}_{p}$ are independent of one another, we get that all
functions $\Psi^{(l)}_{p=1,\ldots,2l+1}$ are eigenfunctions of the
Schr\"{o}dinger equation with energy $E$.  Therefore, such a state
is $(2l+1)$-fold degenerate. We note that an accidental degeneracy
$E_{0}=E_{j\neq 0}$  is impossible because, in this case, all states
of the system would be degenerate, which contradicts the theorem on
the nondegeneracy of GS. Thus, only the state with $L= 0$ is
non-degenerate. This state transits into itself at any rotation and
transforms according to the identity representation of the group
$SO(3)$.

We remark that the first way of the proof follows, in fact, from the
second (group) way since the operators $\hat{L}_{\xi}$ ($\xi=x,y,z$)
are connected with the generators $\hat{I}_{\xi}$  of the group
$SO(3)$ by the relation $\hat{L}_{\xi}=i\hbar\hat{I}_{\xi}$
\cite{elliott,petrashen}. The group analysis is the most general and
establishes the connection between the symmetry of a boundary-value
problem and the properties of wave functions.

Since GS of the system of spinless bosons is non-degenerate
\cite{gilbert} (see also Appendix below), it must correspond to a
state with $L= 0$. This state is isotropic in the sense that WF does
not change under any rotation of all $\textbf{r}_{j}$ around the
center of the system. In this case, the atomic arrangement relative
to each other may be anisotropic and correspond to a crystal, for
instance. Such an anisotropy manifests itself in the two-particle
distribution function and higher ones. That is, \textit{a state with
$L= 0$ may correspond both to a liquid and a crystal} (in Section 5,
this question is considered for a 2D system in greater detail). At
first, the author of the present article was unconscious that the
condition $L= 0$ is compatible with a crystalline solution and
arrived at the wrong conclusion \cite{mtwrong} that the symmetry
analysis forbids the crystalline GS with a nodeless WF.

% pass over

All distribution functions
$g_{j}(\textbf{r}_{1},\ldots,\textbf{r}_{j})$ of a ball-shaped
system are invariant under rotations for any pure state of the
system. Consider, for example, the two-particle distribution
function for a $j$-th state:
\begin{equation}
g_{2}(\textbf{r}_{1},\textbf{r}_{2})=V^{2}\int\limits_{G}
d\textbf{r}_{3}\cdots d\textbf{r}_{N}|\Psi_{j}(\textbf{r}_{1},\ldots
,\textbf{r}_{N})|^{2},
 \label{gF2b}    \end{equation}
where $G=V^{N-2}$ is the domain of the variables
$\textbf{r}_{3},\ldots, \textbf{r}_{N}$. Now express all
$\textbf{r}_{j}$ in the form
$\textbf{r}_{j}=A^{-1}\acute{\textbf{r}}_{j}$, where $A$ denotes the
matrix of rotation around the coordinate origin (center of the
ball). In this case, the Jacobian is $J=det(A^{-1})=1$, and the
boundary surface passes into itself (hence, $V=\acute{V}$ and
$G=\acute{G}$). At any rotation, the state of such a system passes
into an equivalent one. Therefore, the relation
$e^{i\beta}\Psi_{j}(\textbf{r}_{1},\ldots
,\textbf{r}_{N})=\Psi_{j}(\acute{\textbf{r}}_{1},\ldots
,\acute{\textbf{r}}_{N})$ must hold (here, $\beta$ is a constant,
depending on rotation angles; $\beta=0$ for states with $L= 0$). In
the issue we obtain
 \begin{equation}
g_{2}(\textbf{r}_{1},\textbf{r}_{2})=V^{2}\int\limits_{\acute{G}}
d\acute{\textbf{r}}_{3}\cdots
d\acute{\textbf{r}}_{N}|\Psi_{j}(\acute{\textbf{r}}_{1},\ldots
,\acute{\textbf{r}}_{N})|^{2}\equiv
g_{2}(\acute{\textbf{r}}_{1},\acute{\textbf{r}}_{2}),
 \label{gF2c}    \end{equation}
that is, $g_{2}(\textbf{r}_{1},\textbf{r}_{2})$ is isotropic. One
can show similarly that the particle number density $n(\textbf{r})$
and all higher distribution functions are also isotropic under
rotations around the center of the ball.

A crystal is commonly defined as a system that (i) gives the Bragg
peaks when scattering X-rays (the main criterion), (ii) permits the
existence of transverse phonons, and (iii) does not flow (except for
the supersolid state). The Bragg peaks are sharp peaks of the
structure factor $S(\textbf{k})$ which is connected with
$g_{2}(\textbf{r}_{1},\textbf{r}_{2})$ by the equation \cite{vak2}
\begin{equation}
S(\textbf{k})=\langle\rho_{\textbf{k}}\rho_{-\textbf{k}}\rangle =
1+\frac{1}{N}\mathop{\sum_{j,l=1,\ldots,N}}\limits_{j\neq
l}\frac{1}{V^{2}}\int d\textbf{r}_{l}
d\textbf{r}_{j}g_{2}(\textbf{r}_{l},\textbf{r}_{j})e^{-i\textbf{k}(\textbf{r}_{l}-\textbf{r}_{j})}.
 \label{sk1}    \end{equation}
Here, the integration over coordinates $\textbf{r}_{l},
\textbf{r}_{j}$ is to be performed over the volume $V$ of the
system. Since the particles are indistinguishable, we obtain
\begin{equation}
S(\textbf{k})= 1+\frac{N-1}{V^{2}}\int d\textbf{r}_{1}
d\textbf{r}_{2}g_{2}(\textbf{r}_{1},\textbf{r}_{2})e^{-i\textbf{k}(\textbf{r}_{1}-\textbf{r}_{2})}.
 \label{sk2}    \end{equation}
For uniform systems,
$g_{2}(\textbf{r}_{1},\textbf{r}_{2})=g_{2}(\textbf{r}_{1}-\textbf{r}_{2})$.
Then (\ref{sk2}) leads to the well-known formula \cite{balescu}
 \begin{equation}
S(\textbf{k})|_{\textbf{k}\neq 0}=1+n\int
d\textbf{r}(g_{2}(\textbf{r})-1)e^{-i\textbf{k}\textbf{r}},
 \label{gF2a}    \end{equation}
where $g_{2}(\textbf{r})-1$ is the two-particle correlation
function. Our system is finite and generally nonuniform. Therefore,
we will base ourselves on the general formula (\ref{sk2}), which is
valid for a finite system of any form. Let us rotate in Eq.
(\ref{sk2}) the coordinates $\textbf{r}_{l}, \textbf{r}_{j}$
similarly to the above analysis for
$g_{2}(\textbf{r}_{1},\textbf{r}_{2})$ and use the property
$A^{-1}=A^{T}$ \cite{korn}. Then we find
$\textbf{k}\textbf{r}=\textbf{k}(A^{-1}\acute{\textbf{r}})=\textbf{k}(A^{T}\acute{\textbf{r}})=(A\textbf{k})\acute{\textbf{r}}=
\acute{\textbf{k}}\acute{\textbf{r}}$ and
$S(\textbf{k})=S(\acute{\textbf{k}})$. That is, the structure factor
is isotropic and depends only on $|\textbf{k}|$. Consequently, a
state with $L=0$ does not give the Bragg peaks. In this case, the
functions $g_{j\geq 2}(\textbf{r}_{1},\ldots,\textbf{r}_{j})$ may
exhibit a crystal structure with long-range order in the mutual
arrangement of atoms. Under rotation of all coordinates, such a
structure rotates as a whole. As a result, the functions $g_{j\geq
2}(\textbf{r}_{1},\ldots,\textbf{r}_{j})$ do not alter.

\section{Perfectly circular 2D system of spinless bosons}
Let the system be inside a circle of radius $R$ with zero BCs. The
Hamiltonian
 \begin{equation}
 \hat{H} = -\frac{\hbar^{2}}{2m}\sum\limits_{j=1}^{N}\left
[\frac{1}{\rho_{j}}\frac{\partial}{\partial\rho_{j}}\left
(\rho_{j}\frac{\partial}{\partial\rho_{j}} \right )
+\frac{1}{\rho_{j}^{2}}\frac{\partial^{2}}{\partial\varphi_{j}^{2}}\right
] + \frac{1}{2} \mathop{\sum_{j,l}}\limits_{j\neq l}
 U(|\textbf{r}_{l}-\textbf{r}_{j}|)
     \label{u2d} \end{equation}
is invariant under the rotation group  $SO(2)$  and the groups
$T(2)$, $S_{N}$, $C_{i}$. The group $SO(2)$ is compact and abelian,
so all its irreducible representations are one-dimensional
\cite{elliott,petrashen}. In this case, the dimensions of
irreducible representations do not help to ascertain which states
are non-degenerate. One can establish it otherwise. Since
$[\hat{H},\hat{L}_{z}]=0$ (where
$\hat{L}_{z}=-i\hbar\sum_{j=1}^{N}\frac{\partial}{\partial
\varphi_{j}}$ is the operator of the total angular momentum of the
system), the complete set of wave functions can be constructed so
they to be the eigenfunctions of the operators $\hat{H}$ and
$\hat{L}_{z}$. Introduce  the phase inversion operator
$\hat{I}_{\varphi}$: $\varphi_{j}\rightarrow -\varphi_{j}$ for all
$j=1,\ldots,N$.  The states $\Psi(\varphi_{1},\ldots,\varphi_{N})$
and
$\hat{I}_{\varphi}\Psi(\varphi_{1},\ldots,\varphi_{N})=\Psi(-\varphi_{1},\ldots,-\varphi_{N})$
correspond to the same energy $E$ (because
$[\hat{H},\hat{I}_{\varphi}]=0$), but different eigenvalues of the
operator $\hat{L}_{z}$: $L_{z}=\hbar m_{L}$ and $L_{z}=-\hbar
m_{L}$, respectively. Therefore, all states with $m_{L}\neq 0$ are
doubly degenerate. Only states with $m_{L}= 0$ are non-degenerate.
The degeneracy is related here to the fact that
$[\hat{H},\hat{L}_{z}]=0$, $[\hat{H},\hat{I}_{\varphi}]=0$, but
$[\hat{L}_{z},\hat{I}_{\varphi}]=-2\hat{I}_{\varphi}\hat{L}_{z}\neq
0$. Below we find a general form of the wave functions of such a
system and argue that the structure of WFs with $m_{L}= 0$ tolerates
both crystalline and liquid solution.

\subsection{Wave functions for a perfectly circular 2D
system of spinless bosons} For periodic BCs, the wave functions of
weakly excited states of the Bose system and equations for them are
obtained in \cite{holes2020,yuv1,yuv2}. For another BCs, wave
functions of a Bose system and equations for these WFs presumably
have not been obtained (except in the case of a 1D system of point
bosons \cite{gaudin1971,gaudinm,bulatov1988,mtjpa2017}). Let us find
the general structure (without equations) of the WFs of a 2D Bose
system under zero BCs on a circle.

Consider a 2D system of interacting spinless bosons, which is
described by the Hamiltonian (\ref{u2d}) and lies in the circle of
radius $R$. Let BCs be zero ones:
$\Psi(\textbf{r}_{1},\ldots,\textbf{r}_{N})=0$ provided
$\rho_{j}\equiv|\textbf{r}_{j}| =R$ for any $j$. Each eigenfunction
of this boundary-value problem can be expanded in the wave functions
of free bosons (found for the same boundary-value problem) since
they form a complete set of Bose symmetric functions. At a
switched-off interaction, the Schr\"{o}dinger equation reads
\begin{equation}
  -\frac{\hbar^{2}}{2m}\sum\limits_{j=1}^{N}\left [\frac{1}{\rho_{j}}\frac{\partial}{\partial\rho_{j}}\left (\rho_{j}\frac{\partial}{\partial\rho_{j}}
  \right ) +\frac{1}{\rho_{j}^{2}}\frac{\partial^{2}}{\partial\varphi_{j}^{2}}\right ]\psi
  =E\psi.
     \label{shred1} \end{equation}
For free particles $E\geq 0$, so we put
$E=\frac{\hbar^{2}k^{2}}{2m}$ with $k>0$. Then Eq. (\ref{shred1})
takes the form
\begin{equation}
  \sum\limits_{j=1}^{N}\left [\frac{1}{\rho_{j}}\frac{\partial}{\partial\rho_{j}}\left (\rho_{j}\frac{\partial}{\partial\rho_{j}}
  \right ) +\frac{1}{\rho_{j}^{2}}\frac{\partial^{2}}{\partial\varphi_{j}^{2}}\right ]\psi
  +k^{2}\psi=0.
     \label{shred2} \end{equation}
When $N=1$ we set $k\rho=z$ and $\psi(z,\varphi) =
e^{il\varphi}f(z)$. The condition
$\psi(z,\varphi=0)=\psi(z,\varphi=2\pi)$ gives $l=0; \pm 1; \pm 2;
\ldots; \pm \infty$. In the issue at $N=1$, Eq. (\ref{shred2}) leads
to the Bessel equation
\begin{equation}
  z^{2}\frac{d^{2}f}{dz^{2}}+z\frac{df}{dz}+(z^{2}-l^{2})f=0.
     \label{shred3} \end{equation}
Its general solution is $f(z)=c_{j}J_{l}(z)+c_{y}Y_{l}(z)$, where
$J_{l}(z)$ and $Y_{l}(z)$ are the Bessel function  and the Weber
function, respectively \cite{whittaker,tih,jahnke}. In view of the
relations $J_{-l}(z)=(-1)^{l}J_{l}(z)$ and
$Y_{-l}(z)=(-1)^{l}Y_{l}(z)$ \cite{jahnke} one can write
$f(z)=c_{j}J_{|l|}(z)+c_{y}Y_{|l|}(z)$. Since $Y_{|l|}(z\rightarrow
0)\rightarrow \infty$, we reject the Weber function. Thus, Eq.
(\ref{shred2}) with $N=1$ has the solutions $\psi_{l,k} =
e^{il\varphi}J_{|l|}(k\rho)$ ($l=0; \pm 1; \ldots; \pm \infty$) with
the boundary condition $\psi_{l,k}(\rho =R)=0$. Similarly,
separating variables, we find the solutions of Eq. (\ref{shred2})
for $N>1$ particles:
\begin{equation}
 \psi_{\{l_{j}\},\{k_{j}\}}(\textbf{r}_1,\ldots,\textbf{r}_N) =c_{\{l_{j}\},\{k_{j}\}}\sum\limits_{P}e^{il_{1}\varphi_{1}+\ldots
 +il_{N}\varphi_{N}}
\prod\limits_{j=1}^{N}J_{|l_{j}|}(k_{j}\rho_{j}),
     \label{wf1-2d} \end{equation}
where $\sum_{j=1}^{N}k_{j}^{2}=k^{2}$,  $\sum_{j=1}^{N}l_{j}=l_{z}$,
$c_{\{l_{j}\},\{k_{j}\}}$ is a normalization constant, $\sum_{P}$
denotes the sum over all possible permutations of coordinates
$(\varphi_{1},\rho_{1}),\ldots,(\varphi_{N},\rho_{N})$. Here
$l_{z}=L_{z}/\hbar$ is fixed; it is the quantum number of the total
angular momentum of the system. The sets
$\{k_{j}\}=(k_{1},\ldots,k_{N})$ for each set $\{l_{j}\}$ must be
found from the BC $J_{|l_{1}|}(k_{1}R)=\ldots =
J_{|l_{N}|}(k_{N}R)=0$ (there is no need to take the equation
$\sum_{j=1}^{N}k_{j}^{2}=k^{2}$ into account,
% the equation .. needs not be taken into account, since
since $k^{2}$ is arbitrary). Each set $\{l_{j}\}$ must correspond to
an infinite number of sets $\{k_{j}\}$ (because $J_{l}(z)|_{|z|\gg
1,|l|}\approx \frac{\sqrt{2}}{\sqrt{\pi
z}}\cos{(z-\frac{l\pi}{2}-\frac{\pi}{4})}$
\cite{whittaker,tih,jahnke} and $z=k_{j}\rho\leq k_{j}R$ takes
values on the segment $[0,\infty]$ since $k_{j}\in ]0,\infty]$). The
functions (\ref{wf1-2d}) form a complete set of  orthonormal
functions. The general solution corresponding to a given $l_{z}$ is
a superposition of solutions (\ref{wf1-2d}) with all possible
$\{l_{j}\}=(l_{1},\ldots,l_{N})$, satisfying the condition
$\sum_{j=1}^{N}l_{j}=l_{z}$, and all $\{k_{j}\}$ for each set
$\{l_{j}\}$:
\begin{equation}
\psi(\textbf{r}_1,\ldots,\textbf{r}_N)=
 \mathop{\sum_{\{l_{j}\}}}\limits_{l_{1}+\ldots+l_{N}=l_{z}}
\sum\limits_{\{k_{j}\}}
b(l_{1},\ldots,l_{N};k_{1},\ldots,k_{N})\psi_{\{l_{j}\},\{k_{j}\}}(\textbf{r}_1,\ldots,\textbf{r}_N),
    \label{wf2-2d} \end{equation}
where $l_{j}=0; \pm 1; \ldots; \pm \infty$, $j=1,\ldots,N$, and
$b(l_{1},\ldots,l_{N};k_{1},\ldots,k_{N})$ are constants. The
boundary condition  $J_{|l_{1}|}(k_{1}R)=\ldots =
J_{|l_{N}|}(k_{N}R)=0$ must be satisfied for each function
(\ref{wf1-2d}).

As the two-particle distribution function is isotropic, for each
state it must have the form
$g_{2}(\textbf{r}_{1},\textbf{r}_{2})=f_{1}(\rho_{1},\rho_{2})+\Phi(\varphi_{1}-\varphi_{2})f_{2}(\rho_{1},\rho_{2})$,
where $f_{1}$, $f_{2}$, and $\Phi$ are some functions. It is
essential that $\Phi(\varphi_{1}-\varphi_{2})\neq const$, otherwise
the dependence on angles would drop out, which is unphysical: such
dependence is always present in real-life systems because the
correlations between two atoms cannot depend only on $\rho_{1}$ and
$\rho_{2}$. As is seen from the properties of the function
$e^{il\varphi}$ and the Bessel functions, if in the total WF
(\ref{wf2-2d}) the main weight falls on terms with
$|l_{1}|,\ldots,|l_{N}|\sim 1$ then the solution should describe a
liquid. A crystalline solution with any $l_{z}$ (including
$l_{z}=0$) may correspond to WF (\ref{wf2-2d}) with the main
contribution from harmonics corresponding to  $k_{j}\simeq 2\pi/a$
and $|l_{j}|$ of various magnitude from $\sim 1$ to $\sim \sqrt{N}$
(here, $a$ denotes the crystal lattice spacing).

States with $l_{z}\neq 0$ are degenerate and therefore must be
excited states of the system. For such states, the total WF
(\ref{wf2-2d}) can be written in the form $\psi=\psi_{q}\psi_{0}$,
where $\psi_{0}$ is a nodeless WF of the ground state of the system
and $\psi_{q}$ describes a single quasiparticle or a set of
interacting quasiparticles. In this case, GS is that of states with
$l_{z}= 0$, which has the lowest energy. We may expect from physical
considerations that, at weak coupling, the function $\psi_{q}$ for a
single quasiparticle is approximately reduced to the solution for a
free particle: $\psi_{q}=\psi_{l_{z}k_{q}}\approx
\frac{c_{l_{z},k_{q}}}{\sqrt{N}}\sum_{j}
e^{il_{z}\varphi_{j}}J_{|l_{z}|}(k_{q}\rho_{j})$ \cite{cr1}.

 \section{Conclusions}
We have found that the perfect symmetry of a boundary-value problem
can lead to unconventional properties  of the system. In particular,
the density $\rho(\textbf{r})=mn(\textbf{r})$ of a periodic system
turns out to be a constant, and the two-particle distribution
function of a system with rotational symmetry is always isotropic
(therefore, a scattering will not produce the Bragg peaks). That is,
perfect symmetry hides those inner properties of the system, which
``break'' this symmetry. In the examples above, these are
nonuniformity and anisotropy. In real-life 1D and 2D systems,
external fields necessarily violate the translation invariance of
the Hamiltonian. While a real 3D system can be periodic only if it
is the entire Universe. Therefore in realizable systems,
$\rho(\textbf{r})$ should not be a constant. Similarly with
rotational symmetry: for real systems, the ideal rotational symmetry
of BCs is at least slightly broken. In this case, a rotation does
not transfer the system into an equivalent state, and the task
becomes similar to that with strongly broken rotational symmetry. As
a consequence, for the crystalline state, the two-particle
distribution function should be anisotropic, and the Bragg peaks
must be observed.

Another result is that symmetry together with Courant-Hilbert's
theorem narrows the possible properties of the ground state vastly
but does not enable a choice to be made between crystalline and
liquid ground state: both are equally permissible. We remark that
the symmetry analysis cannot be used for real-life BCs, not having
ideal symmetry. Because the symmetry analysis requires that both the
Hamiltonian and the boundary conditions have the same symmetry.

The symmetry analysis allows one to ascertain an interesting
property: the liquid to crystal transition does \textit{not} imply
spontaneous symmetry breaking (SSB). Indeed, the term ``spontaneous
symmetry breaking'' is applied when the system transits from the
state possessing the symmetry of the Hamiltonian to a state with
lower symmetry. However, for the above-considered systems, the
crystalline and liquid states are invariant under all symmetry
transformations of the Hamiltonian and BCs. In this case, the liquid
to crystal transition  implies a physical change in symmetry since
such a transition leads to a change in the properties of
distribution functions. In real systems, the symmetry of BCs is much
lower than the symmetry of the Hamiltonian. Therefore, crystalline
and liquid solutions at $T=0$ also have lower symmetry, which
corresponds to boundary conditions. In this case, we virtually
cannot verify whether SSB occurs under the given phase transition.
So it seems correct to associate the liquid-to-crystal transition
with spontaneous symmetry \textit{reduction}, distinguishing the
latter from SSB. Reduction of symmetry at the liquid-to-crystal
transition can be ascertained by those changes in the properties of
thermodynamic quantities and the structure factor that are
associated with changes in the behavior of the distribution
functions $g_{j\geq 2}(\textbf{r}_{1},\ldots,\textbf{r}_{j})$. All
these properties evidence that the crystal solution is invariant
under some discrete rotation group being a subgroup of the
continuous group $O(3)=SO(3)\times C_{i}$, but the liquid solution
is invariant under the group $O(3)$ itself.

The author wishes to express his gratitude to V.~Gusynin,
S.~Sokolov, and Yu.~Shtanov for discussions and valuable remarks. I
am also grateful to the anonymous referee for helpful comments. This
research is supported in part by the National Academy of Sciences of
Ukraine (project No.~0121U109612).

% для ФНТ нужно добавить: в конце статьи Автор, Абстракт и ключевые слова на укр;
% остался вопрос, как правильно писать crystalline
% and liquid solution или a crystalline and a liquid solution
% здесь предполагается, что это некое крист или жидк решение,
% точно не известное. Correct: crystalline and liquid solution
% Еще вопрос физический: реально всегда система находится в неком  внешнем поле,
% электрическом и магнитном. Не приведет ли это для 2Д случая к решениям с статистикой? !! Иследовать

\section{Appendix. Proof of the nondegeneracy of the ground state}
Consider the theorem on the nondegeneracy of GS and constraints for
this theorem. Although the theorem is well-known, its proof is less
known, and the conditions for the applicability of this theorem were
seemingly not discussed in the literature.

In the classical monograph by R. Courant and D. Hilbert
\cite{gilbert}, the node theorem has been proved for one spinless
particle located in a finite two-dimensional volume with zero BCs.
The proof can be easily generalized to the case of a large number of
particles and any dimensionality of space. The proof in
\cite{gilbert} admits the presence of a finite degeneracy
($E_{j}=E_{j-1}$; the degeneracy is finite if the volume of the
system is finite; see \cite{gilbert}, $\S 2$). If GS is doubly
degenerate, then one of the states is described by a nodeless WF
$\psi_{1}$ (according to the node theorem). WF $\psi_{2}$ of the
second state can have a single node, according to the same theorem.
On the other hand, $\psi_{2}$ should be orthogonal to $\psi_{1}$
and, therefore, \textit{must} have at least one node. Thus, the node
theorem \cite{gilbert} admits a degeneracy of the ground state.
% the possibility for GS is to be degenerate.
%the possibility of GS is degenerate.

The nondegeneracy of GS has been proved at the other place of book
\cite{gilbert}. The proof is based on Jacobi's method (see
\cite{gilbert}, $\S 7$). We will give it in a slightly more detailed
form.

Consider the Schr\"{o}dinger equation
\begin{eqnarray}
   -\triangle \psi + U(x,y)\psi - E \psi=0
       \label{s1}\end{eqnarray}
for one particle located in a 2D region $G=(x,y)$ with zero BCs
($\psi(x,y)=0$ on the boundary of the region $G$). Here, $U(x,y)$ is
a potential, and we set $\hbar=2m=1$. If there exists a solution
$\psi_{1}$ of Eq. (\ref{s1}) corresponding to the smallest
eigenvalue $E_{1}$, then $\psi_{1}$ can be found by solving the
following variation problem \cite{gilbert}: the inequality
\begin{eqnarray}
 D[\varphi]=\int\limits_{G}dx dy (\varphi_{x}^{2}+\varphi_{y}^{2}+U\varphi^{2})\geq  E_{1}\int\limits_{G}dx dy \varphi^{2}
       \label{s2}\end{eqnarray}
must be satisfied for all functions $\varphi(x,y)$ that are equal to
zero on the boundary of the region $G$ and have ``good'' properties
($\varphi$ is to be continuous, whereas $\varphi_{x}$ and
$\varphi_{y}$ are to be piecewise continuous). Here,
$\varphi_{x}\equiv
\partial\varphi/\partial x$, $\varphi_{y}\equiv
\partial\varphi/\partial y$.  Inequality (\ref{s2})
becomes  equality only for $\varphi(x,y)=c_{1}\psi_{1}(x,y)$, where
$c_{1}=const$. It follows from the node theorem that $\psi_{1}$ has
no nodes \cite{gilbert}.

Assume that GS is degenerate and corresponds to two functions:
$\psi_{1}$ and $\psi_{2}$. In this case, $\psi_{1}$ has no nodes,
and $\psi_{2}$ must have one node (as was noted above). In this
case, from the variation viewpoint, $\psi_{1}$ and $\psi_{2}$
satisfy condition (\ref{s2}) and the zero BCs, and $\psi_{2}$
additionally satisfies the condition of orthogonality of the
functions $\psi_{1}$ and $\psi_{2}$. Since $\psi_{1}$ has a constant
sign everywhere inside $G$, we may set
$\psi_{2}(x,y)=\vartheta(x,y)\psi_{1}(x,y)$. Let us ascertain
whether such a solution is possible. We set
$\varphi(x,y)=\eta(x,y)\psi_{1}(x,y)$ in $D[\varphi]$ (\ref{s2}).
Then
\begin{eqnarray}
 D[\varphi]=\int\limits_{G}dx dy [\psi_{1}^{2}(\eta_{x}^{2}+\eta_{y}^{2})+\eta^{2}(\psi_{1x}^{2}+\psi_{1y}^{2})+
 2\psi_{1}\psi_{1x}\eta\eta_{x}+
 2\psi_{1}\psi_{1y}\eta\eta_{y}+U\eta^{2}\psi_{1}^{2}].
       \label{s3}\end{eqnarray}
Let us use the relations  $2\eta\eta_{x}= (\eta^{2})_{x}$,
$2\eta\eta_{y}= (\eta^{2})_{y}$ and integrate the terms with
$\eta\eta_{x}$ and $\eta\eta_{y}$ by parts. We obtain two integrals
over the boundary, both are zero due to the zero BCs, and the
remaining terms give
\begin{eqnarray}
 D[\varphi]=\int\limits_{G}dx dy [\psi_{1}^{2}(\eta_{x}^{2}+\eta_{y}^{2})-\eta^{2}\psi_{1}\triangle \psi_{1}+
 U\eta^{2}\psi_{1}^{2}].
       \label{s4}\end{eqnarray}
Since $\psi_{1}$ satisfies Eq. (\ref{s1}) with $E=E_{1}$, formula
(\ref{s4}) is reduced to
\begin{eqnarray}
 D[\varphi]=\int\limits_{G}dx dy [\psi_{1}^{2}(\eta_{x}^{2}+\eta_{y}^{2})+
 E_{1}\eta^{2}\psi_{1}^{2}]\geq E_{1}\int\limits_{G}dx dy \varphi^{2}.
       \label{s5}\end{eqnarray}
The equality is obtained only for $\eta(x,y)=C=const$. Hence, the
wave function corresponding to the energy $E_{1}$ can have only the
form $const\cdot\psi_{1}(x,y)$. Therefore, the solution
$\psi_{2}(x,y)=\vartheta(x,y)\psi_{1}(x,y)$ with $\vartheta(x,y)\neq
const$ is impossible. It proves that the lowest level is
indispensably non-degenerate. In this case, any excited state
$\psi_{j>1}(x,y)$ can be degenerate (because $\psi_{j>1}(x,y)$ has
nodes and, therefore, the representation
$\varphi(x,y)=\eta(x,y)\psi_{j}(x,y)$ is inapplicable).

If we pass from $x,y$ to $\textbf{r}_{1},\ldots,\textbf{r}_{N}$ in
all formulae, the reasoning remains valid. Therefore, the conclusion
about the nondegeneracy of GS is valid for systems with any $N$ and
for any dimensionality of space. The above analysis is carried out
for a system under the zero BCs. We may expect that the main
conclusion holds under any BCs.

An important point is the conditions under which this theorem works:
(1) wave functions are real single-component; (2) the particle
system is finite; (3) the particles do not have a spin, intrinsic
dipole and higher multipole moment; (4) the total potential energy
is finite: $-\infty < \int d\textbf{r}_{1}\cdots d\textbf{r}_{N}
U(\textbf{r}_{1},\ldots,\textbf{r}_{N})|\Psi_{0}(\textbf{r}_{1},\ldots
,\textbf{r}_{N})|^{2}< \infty$. Conditions (1) and (2) are assumed
in the proof. The violation of condition (4) makes impossible the
transformation of inequality (\ref{s2}) into equality at finite
$E_{1}$. Particles should be spinless because spin indirectly
influences the structure of spatial WFs and energy levels of the
system, which is not taken into account in the theorems by Courant
and Hilbert. Particles cannot possess intrinsic multipole moment
since the node theorem implies that the potential $U$ depends solely
on the spatial coordinates of the particles. More accurately: the
theorem \textit{works} if each particle has a moment whose magnitude
and direction are \textit{fixed}. However, in real systems of
dimensionality $\geq 2$ the directions of the moments can change,
which must lead to an additional strong degeneracy of states. If
particles have the dipole or higher multipole moment, induced by
interaction with neighboring particles, the theorem remains valid
(since such a moment is expressed in terms of the spatial
coordinates of the particles \cite{wb1,wb2,lt2011}).

We note that Courant-Hilbert's theorem on the nondegeneracy of GS
has been proved only for the Schr\"{o}dinger equation (\ref{s1}).
This theorem also holds if the system is placed in an external field
if it appears as a part of the potential $U(x,y)$. But if the
external fields lead to derivatives in the Schr\"{o}dinger equation
that are not reduced to $const \cdot\triangle$, then the proof above
does not work, and GS may be degenerate. In particular, the
degeneracy of the lowest Landau level \cite{land3,vak} is related to
the derivatives $\partial/\partial x$ and $\partial/\partial y$ (or
$\partial/\partial \varphi$) in the Schr\"{o}dinger equation.
Therefore, this degeneracy does not contradict to Courant-Hilbert's
theorem. WFs with $\theta$ symmetry also arise for the
Schr\"{o}dinger equation containing similar derivatives
\cite{khare}. In both these cases, the external field is magnetic.

As another example, consider a ferromagnetic. Its GS corresponds to
all codirectional atomic spins.
%, directed in the same direction.
In this case, spontaneous symmetry breaking occurs, and GS is
infinitely degenerate under rotations, which contradicts
Courant--Hilbert's theorem. Low-lying levels of such a system are
described by the exchange Hamiltonian \cite{akhiezer}
\begin{eqnarray}
 \hat{H}=-\frac{1}{2}\mathop{\sum_{l,j}}\limits_{l\neq j}J(\textbf{R}_{l}-\textbf{R}_{j})\hat{\textbf{s}}_{l}\hat{\textbf{s}}_{j},
       \label{s6}\end{eqnarray}
where $\textbf{R}_{l}$ denote the coordinates of lattice sites. In
this instance, the violation of Courant--Hilbert's theorem is
related to the use of the spin operators $\hat{\textbf{s}}_{l}$ (the
potential $U$ in (\ref{s1}) is an ordinary function). However,
formally the Hamiltonian (\ref{s6}) follows from an ordinary
Hamiltonian, which describes interacting  nuclei and electrons, and
contains no spin operators. Nevertheless, spin properties are
present implicitly because the exchange interaction is a principal
one only for atoms with the electronic shell of a certain structure.
To comprehend this structure, one needs to take account of the Pauli
principle and other properties of particles with a spin. In this
instance, Courant--Hilbert's theorem is violated because the
particles have a spin. The theorem works exclusively for spinless
particles.

% sections are rearranged; a discussion of fractional statistics is added in section 2


\begin{thebibliography}{200}
\bibitem {ll1963}  E.H. Lieb,  W.~Liniger, \textit{Phys. Rev.} \textbf{130}, 1605 (1963).
 https://doi.org/10.1103/PhysRev.130.1605

\bibitem {gaudin1971}  M. Gaudin, \textit{Phys. Rev. A}  \textbf{4}, 386 (1971).
 https://doi.org/10.1103/PhysRevA.4.386

\bibitem {gaudinm} M. Gaudin, {\it The Bethe Wavefunction}, Cambridge University Press, Cambridge (2014).
 https://doi.org/10.1017/CBO9781107053885

\bibitem {mt2022}  M. Tomchenko,  \textit{J. Phys. A: Math. Theor.} \textbf{55}, 135203 (2022).
 https://doi.org/10.1088/1751-8121/ac552b

\bibitem {gilbert} R. Courant, D. Hilbert, {\it Methods of Mathematical Physics}, Vol.
1, John Wiley \& Sons, New York (1989),  Chapt. VI.

\bibitem {fastovskii1967} V.G. Fastovskii, A.E. Rovinskii, Yu.V. Petrovskii, {\it Inert
gases}, Israel Program for Scientific Translations, Jerusalem
(1967).

\bibitem {audi2003}   G.~Audi, O.~Bersillon, J.~Blachot, A.H.~Wapstra,  \textsl{Nucl. Phys.~A} \textbf{729}, 3 (2003).
 https://doi.org/10.1016/j.nuclphysa.2003.11.001

\bibitem {elliott} J.P. Elliott, P.G. Dawber, {\it Symmetry in Physics}, vol. 1, 2, Macmillan Press, London (1979).

\bibitem {courant}  R. Courant, \textit{Nachr. Ges. Wiss. G\"{o}ttingen. Math.-phys. Kl.}, 81
(1923).  [To see the original publication, visit
https://gdz.sub.uni-goettingen.de/id/PPN252457811\_1923]

\bibitem {aziz1984}  R.A. Aziz,  {\it Interatomic Potentials for Rare-Gases: Pure and Mixed Interactions}.
In: {\it Inert Gases: Potentials, Dynamics, and Energy Transfer in
Doped Crystals}, ed. by M.L. Klein, Springer series in chemical
physics, v. 34, pp. 5--86; Springer-Verlag, Berlin (1984).
https://doi.org/10.1007/978-3-642-82221-6

\bibitem {aziz1991}   R.A.~Aziz, M.J. Slaman \textsl{J.~Chem. Phys.} \textbf{94}, 8047 (1991).
 http://doi.org/10.1063/1.460139

\bibitem {rovenchak2000}   I.O. Vakarchuk, V.V. Babin, A.A. Rovenchak, \textsl{J. Phys. Stud.} \textbf{4}, 16 (2000).
 https://doi.org/10.30970/jps.04.16

\bibitem {mt2005} M.D. Tomchenko,  \textit{Ukr. J. Phys.} \textbf{50}, 720
   (2005).  http://archive.ujp.bitp.kiev.ua/files/journals/50/7/500717p.pdf

\bibitem {petrashen} M.I.  Petrashen, E.D. Trifonov, {\it
Applications of Group Theory in Quantum Mechanics}, Dover
Publications, Mineola, New York (2013).

\bibitem {leinaas1977}  J.M. Leinaas, J. Myrheim, \textit{Nuovo Cimento B} \textbf{37}, 1 (1977).
 https://doi.org/10.1007/BF02727953

\bibitem {leinaas1991}  J.M. Leinaas, J. Myrheim, \textit{Int. J. Mod. Phys. B} \textbf{5}, 2573 (1991).
 https://doi.org/10.1142/S0217979291001024

\bibitem {khare} A. Khare,  \textit{Fractional Statistics and Quantum
Theory}, World Scientific, Singapore (2005).
https://doi.org/10.1142/5752

\bibitem {wu1984}  Y.S. Wu, \textit{Phys. Rev. Lett.} \textbf{52}, 2103 (1984).
 https://doi.org/10.1103/PhysRevLett.52.2103

\bibitem {haldane1991}  F.D.M. Haldane, \textit{Phys. Rev. Lett.} \textbf{67}, 937 (1991).
 https://doi.org/10.1103/PhysRevLett.67.937

\bibitem {murthy1994a}  M.V.N. Murthy, R. Shankar, \textit{Phys. Rev. Lett.} \textbf{72}, 3629 (1994).
 https://doi.org/10.1103/PhysRevLett.72.3629

\bibitem {wu1994a}  Y.S. Wu, \textit{Phys. Rev. Lett.} \textbf{73}, 922 (1994).
 https://doi.org/10.1103/PhysRevLett.73.922

\bibitem {krive1987} I.V. Krive,  A.S. Rozhavskii, \textit{Sov. Phys. Usp.} \textbf{30},
370 (1987). http://doi.org/10.1070/PU1987v030n05ABEH002884

\bibitem {laughlin1999}  R.B. Laughlin, \textit{Rev. Mod. Phys.} \textbf{71}, 863 (1999).
 https://doi.org/10.1103/RevModPhys.71.863

\bibitem {bernard1994}  D. Bernard, Y.S. Wu, {\it A Note on Statistical Interactions and the Thermodynamic Bethe Ansatz}.
In: {\it New Developments of Integrable Systems and Long-ranged
Interaction Models}, ed. by M.L. Ge, Y.S. Wu, pp. 10--20; World
Scientific, Singapore (1995). arXiv:cond-mat/9404025;
https://doi.org/10.48550/arXiv.cond-mat/9404025

\bibitem {isakov1994}  S.B. Isakov, \textit{Phys. Rev. Lett.} \textbf{73}, 2150 (1994).
 https://doi.org/10.1103/PhysRevLett.73.2150

\bibitem {holes2020}  M. Tomchenko, \textit{J. Low Temp. Phys.} \textbf{201}, 463
(2020).   https://doi.org/10.1007/s10909-020-02498-z

\bibitem {mt2015}  M. Tomchenko,  \textit{J.~Phys.~A: Math. Theor.} \textbf{48}, 365003
(2015).   https://doi.org/10.1088/1751-8113/48/36/365003

\bibitem {yangs1969}  C.N. Yang, C.P. Yang, \textit{J. Math. Phys.} (N.Y.) \textbf{10}, 1115 (1969).
  https://doi.org/10.1063/1.1664947

\bibitem {mt-therm}  M. Tomchenko, \textit{J. Low Temp. Phys.} \textbf{187}, 251 (2017).
 https://doi.org/10.1007/s10909-017-1738-6

\bibitem {mtsp2019}  M.D. Tomchenko, \textit{Dopov. Nac. Akad. Nauk Ukr.} No. 12, 49
(2019).    https://doi.org/10.15407/dopovidi2019.12.049

\bibitem {polych1989}  A.P. Polychronakos, \textit{Nucl. Phys.~B} \textbf{324}, 597 (1989).
 https://doi.org/10.1016/0550-3213(89)90522-1

\bibitem {murthy1994b}  M.V.N. Murthy, R. Shankar, \textit{Phys. Rev. Lett.} \textbf{73}, 3331 (1994).
 https://doi.org/10.1103/PhysRevLett.73.3331

\bibitem {axilrod1943} B.M. Axilrod, E. Teller, \textit{J. Chem. Phys.} \textbf{11}, 299 (1943).
 http://doi.org/10.1063/1.1723844

\bibitem {bruch1973} L.W. Bruch, I.J. McGee, \textit{J. Chem. Phys.} \textbf{59}, 409 (1973).
 http://doi.org/10.1063/1.1679820

\bibitem {loubeyre1988}  P. Loubeyre, \textit{Phys. Rev. B}  \textbf{37}, 5432 (1988).
 https://doi.org/10.1103/PhysRevB.37.5432

\bibitem {boronat1994}  J. Boronat, J. Casulleras, \textit{Phys. Rev. B}  \textbf{49}, 8920 (1994).
 https://doi.org/10.1103/PhysRevB.49.8920

\bibitem {gross1962} E.P. Gross,  \textit{Ann. Phys.} \textbf{20}, 44 (1962).
 https://doi.org/10.1016/0003-4916(62)90115-X

\bibitem {woo1972} C.-W. Woo, \textit{Phys. Rev. A} \textbf{6}, 2312 (1972).
 https://doi.org/10.1103/PhysRevA.6.2312

\bibitem {feenberg1974} E. Feenberg,  \textit{Ann. Phys.} \textbf{84}, 128 (1974).
 https://doi.org/10.1016/0003-4916(74)90296-6

\bibitem {yuv1} I.A. Vakarchuk, I.R. Yukhnovskii, \textit{Theor. Math. Phys.} \textbf{40}, 626
(1979).  https://doi.org/10.1007/BF01019246

\bibitem {mcmillan1965}  W.L. McMillan, \textit{Phys. Rev.} \textbf{138}, A442 (1965).
 https://doi.org/10.1103/PhysRev.138.A442

\bibitem {chester1970}  G.V. Chester, \textit{Phys. Rev.} \textbf{A 2}, 256 (1970).
 https://doi.org/10.1103/PhysRevA.2.256

\bibitem {reatto1995} L. Reatto, {\it Boson many-body problem: progress in variational Monte Carlo
computations}, in {\it Progress in Computational Physics of Matter},
ed. by L. Reatto, F. Manghi, pp. 43--98; World Scientific, Singapore
(1995). https://doi.org/10.1142/9789814261319\_0002

\bibitem {whitlock2006}  P.A. Whitlock, S.A. Vitiello,  {\it Quantum Monte Carlo Simulations
of Solid $^{4}$He}. In: {\it Large-Scale Scientific Computing. LSSC
2005}, ed. by I. Lirkov, S. Margenov, J. Wa\'sniewski, Lecture Notes
in Computer Science, vol. 3743, pp. 40--52; Springer, Berlin (2006).
 https://doi.org/10.1007/11666806\_4

\bibitem {nosanow1966} L.H. Nosanow,  \textit{Phys. Rev.} \textbf{146}, 120 (1966).
 https://doi.org/10.1103/PhysRev.146.120

\bibitem {guyer}  R.A. Guyer,  \textit{Solid State Phys.} \textbf{23}, 413 (1970).
 https://doi.org/10.1016/S0081-1947(08)60618-9

\bibitem {cazorla2008}  C. Cazorla, J. Boronat, \textit{J.~Phys. Cond. Mat.} \textbf{20}, 015223 (2008).
 https://doi.org/10.1088/0953-8984/20/01/015223

\bibitem {vak1989} I.A. Vakarchuk, \textsl{Theor. Math. Phys.} \textbf{80}, 983
(1989). https://doi.org/10.1007/BF01016193

\bibitem {vak1990} I.A. Vakarchuk, \textsl{Theor. Math. Phys.} \textbf{82}, 308 (1990).
https://doi.org/10.1007/BF01029225

\bibitem {sacha2018}  K. Sacha, J. Zakrzewski, \textit{Rep. Progr. Phys.} \textbf{81}, 016401 (2018).
 https://doi.org/10.1088/1361-6633/aa8b38

\bibitem {mt2020}  M. Tomchenko, \textit{J. Low Temp. Phys.} \textbf{198}, 100 (2020).
 https://doi.org/10.1007/s10909-019-02252-0

\bibitem {gross1958} E.P. Gross,  \textit{Ann. Phys.} \textbf{4}, 57 (1958).
https://doi.org/10.1016/0003-4916(58)90037-X

\bibitem {gross1960} E.P. Gross,  \textit{Ann. Phys.} \textbf{9}, 292 (1960).
https://doi.org/10.1016/0003-4916(60)90033-6

\bibitem {luca} A. De Luca, L.M. Ricciardi, H. Umezawa, \textit{Physica} \textbf{40}, 61 (1968).
https://doi.org/10.1016/0031-8914(68)90121-3
%  диагонализация гамильтониана --- в работе выше (нечетко) и ниже (аккуратно видимо)

\bibitem {coniglio1969} A. Coniglio, M. Marinaro, B. Preziosi, \textit{Nuovo Cimento B} \textbf{61}, 25 (1969).
https://doi.org/10.1007/BF02711694

\bibitem {kirz} D.A. Kirzhnits, Yu.A. Nepomnyashchi\u{i}, \textit{Sov. Phys. JETP} \textbf{32}, 1191 (1971).
http://jetp.ras.ru/cgi-bin/dn/e\_032\_06\_1191.pdf

\bibitem {nep} Yu.A. Nepomnyashchii, \textit{Theor. Math. Phys.} \textbf{8}, 928 (1971).
https://doi.org/10.1007/BF01029350

\bibitem {shlyapa2015} Z.-K.~Lu, Y.~Li, D.S.~Petrov, G.V.~Shlyapnikov, \textit{Phys. Rev. Lett.} 115, 075303 (2015).
 https://doi.org/10.1103/PhysRevLett.115.075303

\bibitem {andreev2017}  S.V. Andreev, \textit{Phys. Rev. B} \textbf{95}, 184519 (2017).
 https://doi.org/10.1103/PhysRevB.95.184519

\bibitem {fil2020}  D.V.~Fil, S.I.~Shevchenko, \textit{Fiz. Nizk. Temp.} \textbf{46}, 556 (2020) [\textit{Low Temp. Phys.} \textbf{46}, 465 (2020)].
 https://doi.org/10.1063/10.0001049

\bibitem {land3} L.D.~Landau, E.M.~Lifshitz, {\it Quantum Mechanics.
Non-Relativistic Theory}, Pergamon Press, New York (1980).

\bibitem {vak}  I.O. Vakarchuk, {\it Quantum Mechanics}, Lviv University Press, Lviv (2004)
[in Ukrainian].

\bibitem {gelfand} I.M. Gel'fand, R.A. Minlos, Z.Ya. Shapiro, {\it
Representations of the rotation and Lorentz groups and their
applications}, Pergamon Press, New York (1963).

\bibitem {golod} P.I. Holod, A.U. Klimyk, {\it Mathematical Foundations of the  Theory  of
Symmetries}, Kyiv, Naukova Dumka (1992) [in Ukrainian].

\bibitem {mtwrong}  M.D. Tomchenko,  arXiv:2108.03633
[cond-mat.other].  https://doi.org/10.48550/arXiv.2108.03633

\bibitem {vak2}  I.O. Vakarchuk, {\it Introduction into the Many-Body Problem},
Lviv University Press, Lviv (1999) [in Ukrainian].

\bibitem {balescu}  R. Balescu, {\it Equilibrium and Nonequilibrium Statistical
Mechanics}, John Wiley \& Sons, New York (1975).

\bibitem {korn} G.A.~Korn, T.M.~Korn, {\it Mathematical Handbook for Scientists and Engineers:
Definitions, Theorems, and Formulas for Reference and Review},
McGraw-Hill, New York (1968), Chapt. 14.

\bibitem {yuv2} I.A. Vakarchuk,  I.R. Yukhnovskii, \textit{Theor. Math. Phys.} \textbf{42}, 73
(1980). https://doi.org/10.1007/BF01019263

\bibitem {bulatov1988} V.L. Bulatov, \textit{Theor. Math. Phys.} \textbf{75}, 433
(1988).  https://doi.org/10.1007/BF01017178

\bibitem {mtjpa2017}  M. Tomchenko,  \textit{J. Phys. A: Math. Theor.}  \textbf{50}, 055203 (2017).
https://doi.org/10.1088/1751-8121/aa5197

\bibitem {whittaker} E.T. Whittaker,  G.N. Watson, {\it A Course of Modern Analysis},
Cambridge University Press, Cambridge (1996).
https://doi.org/10.1017/CBO9780511608759

\bibitem {tih}   A.N. Tikhonov, A.A. Samarskii, \textit{Equations of Mathematical
Physics}, Dover Publications, New York (2011).

\bibitem {jahnke} E. Jahnke, F. Emde, F. L\"{o}sch,  \textit{Tables of Higher
Functions},  B. G. Teubner Verlagsgesellschaft, Stuttgart (1966).

\bibitem {cr1} V.M.~Loktev, M.D.~Tomchenko, \textit{Ukr. J. Phys.} \textbf{55}, 901
(2010).
http://archive.ujp.bitp.kiev.ua/files/journals/55/8/550807p.pdf

\bibitem {wb1} W. Byers Brown, D.M. Whisnant, \textit{Mol. Phys.} \textbf{25}, 1385
(1973).    https://doi.org/10.1080/00268977300101191

\bibitem {wb2}  D.M. Whisnant, W. Byers Brown, \textit{Mol. Phys.}  \textbf{26}, 1105
(1973).   https://doi.org/10.1080/00268977300102331

\bibitem {lt2011} V.M. Loktev, M.D. Tomchenko, \textit{J.~Phys.~B: At. Mol. Opt. Phys.} \textbf{44}, 035006
 (2011).   https://doi.org/10.1088/0953-4075/44/3/035006

\bibitem {akhiezer}   A.I. Akhiezer, V.G. Bar'yakhtar, S.V. Peletminskii, \textit{Spin
waves}, North-Holland Pub. Co., Amsterdam (1968).
% Pub. Co. --- Publishing Company



\end{thebibliography}
       \end{document}